\documentclass[preprint,prc,showpacs,preprintnumbers,
               superscriptaddress,amsmath,amssymb,floatfix]{revtex4}

\usepackage{graphicx}
\usepackage{dcolumn}
\usepackage{bm}
\usepackage{longtable}

\begin{document}

\title{\large {\bf Nuclear Symmetry Energy in Relativistic
Mean Field Theory  \\ }}

\author{Shufang Ban}
 \affiliation{School of Physics, Peking University,
              Beijing 100871, China}
 \affiliation{Royal Institute of Technology, AlbaNova University Center,
                 10691 Stockholm, Sweden}

\author{Jie Meng}
 \email{mengj@pku.edu.cn}
 \affiliation{School of Physics, Peking University,
              Beijing 100871, China}
 \affiliation{Institute of Theoretical Physics, Chinese Academy of
              Sciences, Beijing 100080, China}
 \affiliation{Center of Theoretical Nuclear Physics, National Laboratory
              of Heavy Ion Accelerator, Lanzhou 730000, China}

\author{  Wojciech Satu{\l}a}
\email{             satula@fuw.edu.pl}
\affiliation{Institute of Theoretical Physics, University of Warsaw,\\
             ul. Ho{\.z}a 69, PL-00 681 Warsaw, Poland}
\affiliation{Royal Institute of Technology, AlbaNova University
Center, 10691 Stockholm, Sweden}

\author{Ramon A. Wyss}
 \affiliation{Royal Institute of Technology, AlbaNova University Center,
                 10691 Stockholm, Sweden}

\date{\today}

\begin{abstract}

The physical origin of the nuclear symmetry energy is studied
within the relativistic mean field (RMF) theory. Based on the
nuclear binding energies calculated with and without mean
isovector potential for several isobaric chains we confirm earlier
Skyrme-Hartree-Fock result that the nuclear symmetry energy
strength depends on the mean level spacing $\varepsilon (A)$ and
an effective mean isovector potential strength $\kappa (A)$. A
detailed analysis of the isospin dependence of the two components
contributing to the nuclear symmetry energy reveals a quadratic
dependence due to the mean-isoscalar potential,
$\sim$\,$\varepsilon\, T^2$, and, completely unexpectedly, the
presence of a strong linear component $\sim$\,$\kappa\, T(T+1+
\varepsilon/\kappa)$ in the isovector potential. The latter
generates a nuclear symmetry energy in RMF theory that is
proportional to $E_{\text{sym}}\sim T(T+1)$ at variance to the
non-relativistic calculation. The origin of the linear term in RMF
theory needs to be further explored.
\end{abstract}


\maketitle


One of the most important topics in current nuclear physics is to
search for the existence limit of atomic nuclei, i.e., to
determine the nuclear drip line.  In this respect, the role of the
continuum in loosely bound nuclei and, in particular, its impact
on the treatment of pairing correlations has been discussed to
great extent in recent time. However, the proper understanding and
correct reproduction of the nuclear symmetry energy (NSE) may have
even greater bearing for masses of loosely bound nuclei and
certainly is a key issue in the study of exotic nuclei. The very
fundamental questions in this respect concern both the
understanding of the microscopic origin of the NSE strength as
well as its isospin dependence. The latter issue has attracted
recently great attention also in $N\sim Z$ nuclei,
see Ref.~\cite{[Sat01x]} and references therein.

The NSE is conventionally parametrized as:
\begin{equation}
   E_{\text{sym}} =  a_{\text{sym}} (A) T(T+\lambda)
\end{equation}
where $T=|T_z| =|N-Z|/2$. The strength of the NSE admits typically
volume and surface components $a_{\text{sym}} (A) = a_{\text{v}}/A
- a_{\text{s}}/A^{4/3}$ and its physical origin is traditionally
explained in terms of the kinetic energy and mean isovector
potential (interaction) contributions i.e. $a_{\text{sym}} (A)a_{\text{kin}} (A) + a_{\text{int}} (A)$,
respectively~\cite{[Boh69]}.  The linear term is found to be
strongly model dependent and there is a common belief that
mean-field models yield essentially only a quadratic term $\lambda
\approx 0$. On the other hand, the nuclear
shell-model~\cite{[Tal56],[Tal62],[Zel96]} or models restoring
isospin symmetry~\cite{[Nee02x]}  suggest that $\lambda \approx 1$.
No consensus is reached so far concerning the value of $\lambda$
although there is certain preference for $\lambda\approx 1$.
Indeed, experimental masses of nuclei with small values of $T$
supports the existence of the linear term~\cite{[Mye66]}. Similar
conclusions were reached by J\"{a}necke {\it et
al.}~\cite{[Jan65x]} based on the analysis of experimental binding
energies for $A<80$ nuclei.
 One of the most accurate mass formula, the so called FRDM ~\cite{[Mol95]}
employs a value of $\lambda\approx 1$ but inconsistently admits
only a volume-like linear term.  Assuming $T(T+1)$ dependence
Duflo and Zuker have performed a global fit to nuclear masses
obtaining~\cite{[Duf95]}
\begin{equation}
 \label{EmpSymmEC}
 a_{\text{sym}} (A)  = \frac{134.4}{A} - \frac{203.6}{A^{4/3}}
                       ~\text{[MeV]}.
\end{equation}

A different view on the origin of the NSE was presented recently
by Satu{\l}a and Wyss. In Refs.~\cite{[Sat03],[Glo04],[Sat05]} it
was demonstrated using the Skyrme-Hartree-Fock (SHF) model that
the NSE can be directly associated with the mean level spacing
$\varepsilon (A)$ and mean isovector potential, $E_{\text{sym}} \dfrac 1 2 \varepsilon (A) T^2 +  \dfrac 1 2 \kappa (A)
T(T+1)$~\cite{[Sat03],[Glo04],[Sat05]}.  Surprisingly, the
self-consistent calculations revealed that the complicated
isovector mean potential induced by the Skyrme force is similar to
that obtained from a simple interaction $\dfrac 1 2 \kappa (A)
\hat{\bf T} \cdot \hat{\bf T}$, i.e., is very accurately
characterized by a single strength $\kappa
(A)$~\cite{[Sat03],[Glo04],[Sat05]}. This study revealed also that
the SHF theory yield in fact a (partial) linear term with $\lambda
\approx \kappa/(2 a_{\text{sym}})$ and that this term originates
from neutron-proton exchange interaction.

Alongside with the SHF calculation, the relativistic mean field
(RMF) theory has been used for a large variety of nuclear
structure phenomena~\cite{[Men96]}. Since the RMF theory is based
on a very different concept from the SHF, it is highly interesting
to investigate the structure of the NSE in the framework of the
RMF theory.

\vspace{0.8cm}

The details of RMF theory together with its applications can be
found in a number of review articles, see for example
Ref.~\cite{[Rin96]} and references therein, and will not be
repeated here. The basic ansatz of the RMF theory is a
Lagrangian density whereby nucleons are described as Dirac
particles which interact via the exchange of various mesons [the
isoscalar-scalar sigma ($\sigma$), isoscalar-vector omega
($\omega$) and isovector-vector rho ($\rho$)] and the photon. The
$\sigma$ and $\omega$ mesons provide the attractive and repulsive
part of the nucleon-nucleon force, respectively. The isospin
asymmetry is provided by the isovector $\rho$ meson. Hence, by
switching on and off the coupling to the
 $\rho$ meson, one can easily separate the role of isoscalar
and isovector parts of the interaction and study them
independently.

In the nuclei considered here, time reversal symmetry is preserved
and the spatial vector components of ${\boldsymbol \omega}$,
$\vec{\boldsymbol \rho}$ and $\mathbf{A}$ fields vanish. This
leaves only the time-like components $\omega^{0}$,
$\vec{\rho}^{\,\,0}$ and $A^{0}$. Charge conservation guarantees
that only the third component of the isovector
$\vec{\rho}^{\,\,0}$ meson is active. For reason of simplicity,
axial symmetry is assumed in the present work. The Dirac
spinor $\psi _{i}$ as well as the meson fields can be expanded in
terms of the eigenfunctions of a deformed axially symmetric
oscillator potential~\cite{[Gam90]} or Woods-Saxon
potential~\cite{[Zho03]}, and the solution of the problem is
transformed into a diagonalization of a Hermitian matrix. 


The RMF calculations are performed for the $A$=40, 48, 56, 88,
100, 120, 140, 160, 164, and 180 isobars with the effective
Lagrangians NL3~\cite{[Lal97]}, TM1~\cite{[Sug95]}, and
PK1~\cite{[Lon04]}. Our choice of the parameterizations
is somewhat arbitrary. However, the purpose of this
work is not to make a detailed comparison to the data but rather
to investigate specific features of the RMF approach  pertaining
to the isovector channel. These properties are expected to be
fairly parameterization independent, in particular that these
parameterizations reproduce rather well the equation of state
for densities $\rho\leq 0.2$\,fm$^{-3}$~\cite{[Ban04],[Men05]}.

The Dirac equations are solved by expansion in the harmonic
oscillator basis with 14 oscillator shells for both the fermion
fields and boson fields. The oscillator frequency of the harmonic
oscillator basis is set to $\hslash \omega _{0}=41A^{-1/3}$ MeV and
the deformation of harmonic oscillator basis $\beta_0$ is reasonably
chosen to obtain the lowest energy. Generally speaking, the RMF
calculation reproduce the experimental binding energy to an accuracy
less than 1\%. For the present study we are mainly interested in the
NSE emerging in the RMF theory due to
the strong (particle-hole) interaction. Hence the Coulomb
potentials and the pairing correlations will be neglected in the
following. The full potential in the Dirac equation is
 \begin{equation}
  V_{tot} = V (\mathbf{r})+ \beta S(\mathbf{r})
          = g_{\omega} \omega^{0}(\mathbf{r})
                + g_{\rho} \vec{\tau} \cdot \vec{\rho}^{\,\,0}(\mathbf{r})
                + \beta g_{\sigma }\sigma(\mathbf{r}).
 \end{equation}
It can easily be separated into isovector and isoscalar components,
i.e.,  $V_{tot}=V_{is}+V_{iv}$, where
\begin{equation}
\begin{array}{lll}
       V_{is}(\mathbf{r})
            & = &  g_{\omega} \omega^{0}(\mathbf{r}) + \beta g_{\sigma}
                      \sigma(\mathbf{r}), \\
       V_{iv}(\mathbf{r})
            & = &
           g_{\rho} \vec{\tau} \cdot \vec{\rho}^{\,\,0}(\mathbf{r}).
\end{array}
\label{vvs}
\end{equation}

 The binding energy calculated with the full potential $V_{tot}$
in Eq.~(3) is denoted as $E_T$. The energy obtained by
switching off the isovector potential, $V_{iv}\equiv 0$, i.e. by taking in
the calculation $V_{tot}\equiv V_{is}$, is denoted
by $\tilde{E}_T$. In order to single out the impact of isoscalar fields
on the NSE, we use $\tilde{E}_T$ to extract the mean level spacing
$\varepsilon(A,T_Z)$ along an isobaric chain
\begin{equation}
      \tilde{E}_T(A,T_z) - \tilde{E}_{T=0}(A,T_z=0)= \dfrac 1 2
          \varepsilon (A,T_z) T^2.
      \label{varepsilonR}
\end{equation}

\begin{figure}[tbp]
\includegraphics[width=7cm,angle=-90]{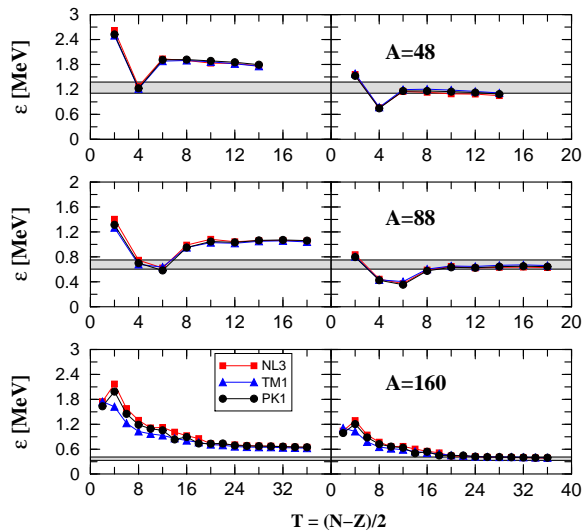}
\caption{\label{fig1}The mean level spacing $\varepsilon$ (left)
          and its counterpart
          (right) scaled by $m^*/m$ for A=48 (upper), 88 (middle) and
          160 (lower) isobaric chains calculated  using
          effective Lagrangians NL3, TM1, and PK1 as marked in the
          figure. The shadowed areas correspond to the empirical
          mean level spacing: $\varepsilon^{emp} = \dfrac {53} {A}
          \sim \dfrac {66} {A} ~\text{MeV}$.}
\end{figure}

The values of  $\varepsilon (A,T_z)$ calculated for the $A$=48,
88, and 160 isobaric chains of nuclei from $T_z=0$ to the vicinity
of the drip line are shown in the left panels of Fig.~\ref{fig1}.
The calculations have been performed for three different
parameterizations of the effective Lagrangian including NL3, TM1,
and  PK1, respectively, which all yield very similar results. For
small values of $T_z$, strong variations in $\varepsilon (A,T_z)$
are seen, which are associated with shell closures. For larger
values of $T_z$, $\varepsilon (A,T_z)$ become less sensitive to
the shell structure and its value is stabilized, $\varepsilon
(A,T_z)\approx \varepsilon (A)$. Note, that the calculated values
of  $\varepsilon (A)$ are much larger then the empirical estimates
for the mean level spacing
 ${53}/{A} \leq \varepsilon^{emp}  \leq {66}/{A}~\text{MeV}$
 ~\cite{[Sat03],[Gil65],[Kat80],[Shl92]}.
However, after rescaling  $\varepsilon^* (A) = ( m^*/
m)\varepsilon (A)$ by the corresponding effective masses
$m^*/m=0.595$, $0.634$, and $0.6055$ for NL3, TM1, and PK1
respectively, the effective mean level spacing $\varepsilon^*(A) $
neatly falls within the empirical bounds (shaded areas), as shown
in the corresponding right panels. Let us also note that with
increasing $A$, i.e., from $A=48$ to $A=160$, all curves move
toward the upper limit of the empirical data, reflecting the
decreasing role of surface effects with increasing $A$, similar to
the SHF results~\cite{[Sat03]}.

When comparing to the results of the SHF calculations in Ref.
\cite{[Sat03]}, the following two important conclusions can be
made: ({\it i\/}) even though the values of $\varepsilon (A)$ from
the RMF calculation are much larger than those from the SHF, after
the effective mass scaling both models generate essentially
identical results in agreement to the empirical boundaries  for
the mean level spacing; ({\it ii\/}) the results from the RMF
calculations clearly confirm the general outcome of
Ref.~\cite{[Sat03]}, that the isoscalar field generates the NSE of
the form~$\varepsilon T^2$ and that its strength is indeed
governed by the mean level density rather than the kinetic energy,
see also~\cite{[Sat05]}. Let us stress that the
contribution~$\varepsilon T^2$ can also be derived analytically
using the simple iso-cranking
model~\cite{[Sat01x],[Sat03],[Glo04]}. Note further, 
that the evidence and
conclusions gathered from the SHF and the RMF calculations are
independent of the iso-cranking model.

 \vspace{0.4cm}

After obtaining the average level density $\varepsilon$, we now
proceed to calculate the average effective strength $\kappa$ of
the isovector potential. The effective isovector potential
strength $\kappa$ is obtained from the binding energy difference
between the RMF calculations with and without the isovector
potential for the same nucleus. As explained later, three
different types of $T$-dependence of the isovector potential are
investigated:
\begin{equation}
      {E_T} - \tilde{E_{T}}= \dfrac 1 2 \kappa T^2 ~~ \text{and} ~~
                             \dfrac 1 2 \kappa T(T+1) ~~ \text{and} ~~
                             \dfrac 1 2 \kappa
                             T(T+1+\varepsilon/\kappa).
      \label{kappaR}
\end{equation}

\begin{figure}[tbp]
\includegraphics[width=7cm,angle=-90]{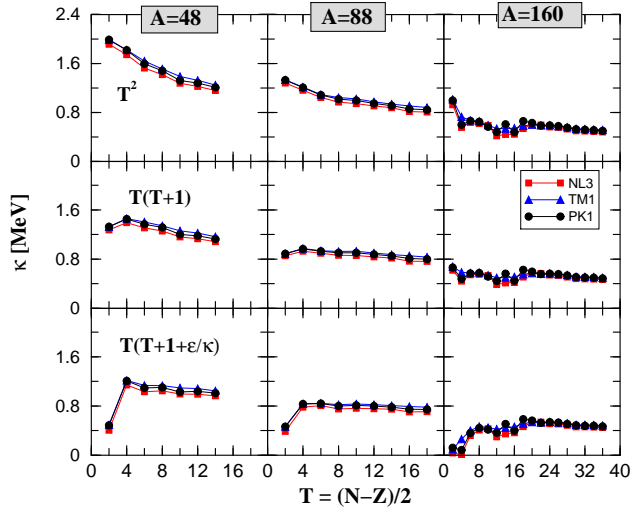}
\caption{\label{fig2} The average effective strength $\kappa$ of
          the isovector potential for A=48 (left), 88 (middle) and
          160 (right) isobaric chains calculated using
          effective Lagrangians NL3, TM1, and PK1 as marked in the
          figure. Upper, middle and lower panels show the values
          of $\kappa$ obtained assuming
          $E_{\text{T}}-{\tilde{E}_{\text{T}}}=\dfrac 1 2 \kappa T^2$,
          $\dfrac 1 2 \kappa T(T+1)$, and
          $\dfrac 1 2 \kappa T(T+1+\varepsilon/\kappa)$, respectively.
          See text for further details.
}
\end{figure}

The resulting effective isovector potential strength $\kappa$ for
A= 48, 88, and 160 isobaric chains in RMF theory are shown in Fig.
\ref{fig2}. Similar to the SHF calculation in Ref. \cite{[Sat03]},
the complicated isovector potential along an isobaric chain can be
characterized by an effective isovector potential strength $\kappa
(A)$.  However, at variance with the SHF~\cite{[Sat03]}, the RMF
dependence on $T$ is best reproduced by a dependence like
      ${E_T} - \tilde{E_{T}} \approx
         \dfrac 1 2 \kappa T(T+1+\varepsilon/\kappa),$
rather than $T^2$ or $T(T+1)$. Apparently, the linear term in RMF
is considerably larger than that in the SHF calculation, implying
that the total NSE in RMF behaves effectively as:
\begin{equation}
      E_{\text{sym}} \approx \dfrac 1 2 \varepsilon T^2 +
                       \dfrac 1 2 \kappa T(T+1+\varepsilon/\kappa)
                     \approx \dfrac 1 2 (\varepsilon+\kappa) T(T+1).
      \label{SymmE}
\end{equation}

The important aspect to note is that in the SHF approximation the
linear term originates predominantly from the Fock exchange in the
isovector channel. In the RMF, which
is a Hartree approximation, one would therefore expect a $\kappa T^2$
dependence.  In contrast, the large slope of the isovector potential
strength $\kappa (A,T_z)$, fitted using either $\kappa T^2$  or
$\kappa T(T+1)$ dependence reveals the presence of an effective
linear term that is large enough to  compensate the lack of the
linear term in the $\varepsilon$ proportional term,
see Fig.~\ref{fig2}. Similar tendencies for the mean level spacing
$\varepsilon$ and effective isovector potential strength $\kappa$
are present in all the 10 isobaric chains we calculated,
including $A$=40, 48, 56, 88, 100, 120, 140, 160, 164, and
180.

At a first glance, the RMF theory is a
Hartree approximation (without exchange term) and one does not expect
a linear term to be present. On the other hand, the RMF as well as the SHF
are two particular realizations of the density functional (DF) theory.
Moreover, due to the large meson masses the relativistic forces
should be close to zero range forces and both approaches are expected to be
rather alike since in this limit the exchange term takes the same form
as the direct term and should be effectively included within the DF.
The question that arises is why the RMF is capable
to generate the linear term in contrast to the SHF approach?
In this context
it is interesting to observe that  within the RMF approach
the isovector mean-potential is generated by the  $\rho$-meson field.
Hence, its properties are defined essentially by a single coupling constant
$g_\rho$, see Eq.~(4). In this respect the RMF seems to
be more flexible than the SHF where isoscalar and isovector parts of the
Skyrme local energy density functional are strongly dependent upon each
other through the auxiliary Skyrme force (SF)
parameters which are fitted to the data. In the process of
detreminig the SF, one should therefore balance properly and very
carefully the isoscalar and isovector data used in the fitting procedure.
Indeed, our earlier study~\cite{[Sat03]} shows that the SkO~\cite{[Rei99]}
parameterization of the SF, which has been fitted to neutron-rich
nuclei, has a stronger linear term than the so called standard
parameterizations but not as strong as that in RMF theory.

 \vspace{0.4cm}

\begin{figure}[tbp]
\includegraphics[width=8cm,angle=-90]{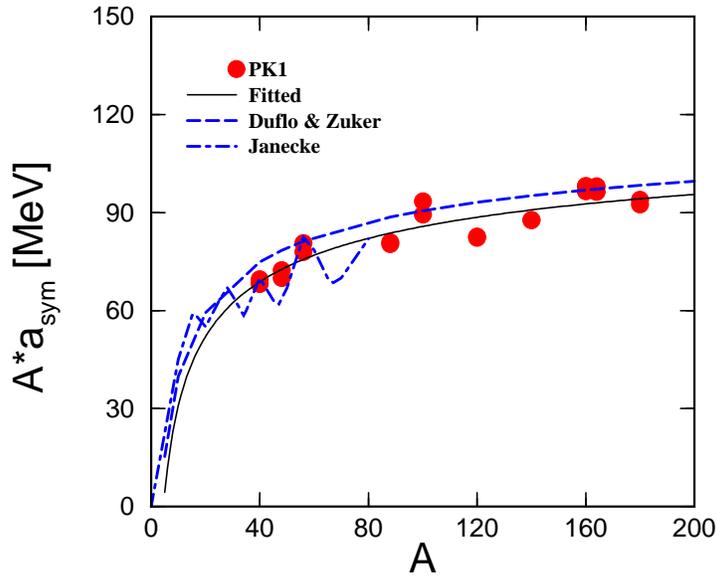}
\caption{\label{fig3new} The nuclear symmetry energy coefficient
          $a_\text{sym}*A$ (filled circles) extracted from
          $E_\text{sym} = a_\text{sym} T(T+1)$ for
          the A=40, 48, 56, 88, 100, 120, 140, 160, 164, and
          180 isobaric chains calculated using
          effective Lagrangian PK1.
          The solid line represents fit
          to the calculations, i.e.,
          $A*a_\text{sym}= 133.20 - 220.27/A^{1/3}$ from Eq. (\ref{SymmEC}).
          The dashed line is from Eq. (\ref{EmpSymmEC}):
          $A*a_\text{sym}= 134.4 - 203.6/A^{1/3}$ \cite{[Duf95]} and
          the dot-dashed line is from Ref.~\cite{[Jan65x]}.
          For further details, see text.}
\end{figure}

We proceed further by investigating the mass dependence of the
NSE. Since the total NSE in RMF theory behaves effectively like
$T(T+1)$ we extract the NSE strength $a_\text{sym}(A)$ from the
difference of the binding energies:
 \begin{equation}
      E_\text{sym}={E_T} - E_{T=0} = a_\text{sym}(A) T(T+1).
      \label{asymR}
 \end{equation}
 To avoid the influence of shell structure we chose two nuclei
 with large $T_z$ for each isobaric chain $A$ and calculate
 $a_\text{sym}(A)$ simply as an arithmetic
 mean over such a pair of nuclei.
 In order to compare with the empirical data of Ref.~\cite{[Duf95]},
 we depict in Fig.~\ref{fig3new} the product
 $A*a_\text{sym}$ as a function of $A$ for all the 10 isobaric
 chains 
 calculated with the effective Lagrangian PK1.
 The dot-dashed line in Fig.~\ref{fig3new} is taken from
 Ref.~\cite{[Jan65x]} revealing the shell structure of the symmetry
 energy coefficient and the dashed line  represents a fit to
 experimental data given by Eq.~(\ref{EmpSymmEC})~\cite{[Duf95]}.
 The maxima at $A$=40 and $A$=56 can easily be seen
in our calculation, which is in good agreement with
Ref.~\cite{[Jan65x]}. This behavior is easy to understand since
$T$=0 nuclei in the $A$=40, 56, 100 and 164 isobaric chains are
double magic nuclei and hence,  more bound, resulting in an
increase of the symmetry energies for nuclei with $T>0$. Still,
the average of the calculated symmetry energy is quite close to
the dashed line, which is fitted to masses in Ref.~\cite{[Duf95]}.

Restricting this analysis to volume and surface terms only, a
least-square fit to the calculated points (filled circles in
Fig.~\ref{fig3new}) leads to a smooth curve:

\begin{equation}
      a_\text{sym}^{(\text{RMF})}
                  = \dfrac {133.20} {A} -
                    \dfrac {220.27} {A^{4/3}} ~\text{[MeV]},
      \label{SymmEC}
\end{equation}
shown as solid line, which is very close to the empirical values
(dashed line).

\vspace{0.4cm}

\begin{figure}[tbp]
\includegraphics[width=6.0cm,angle=-90]{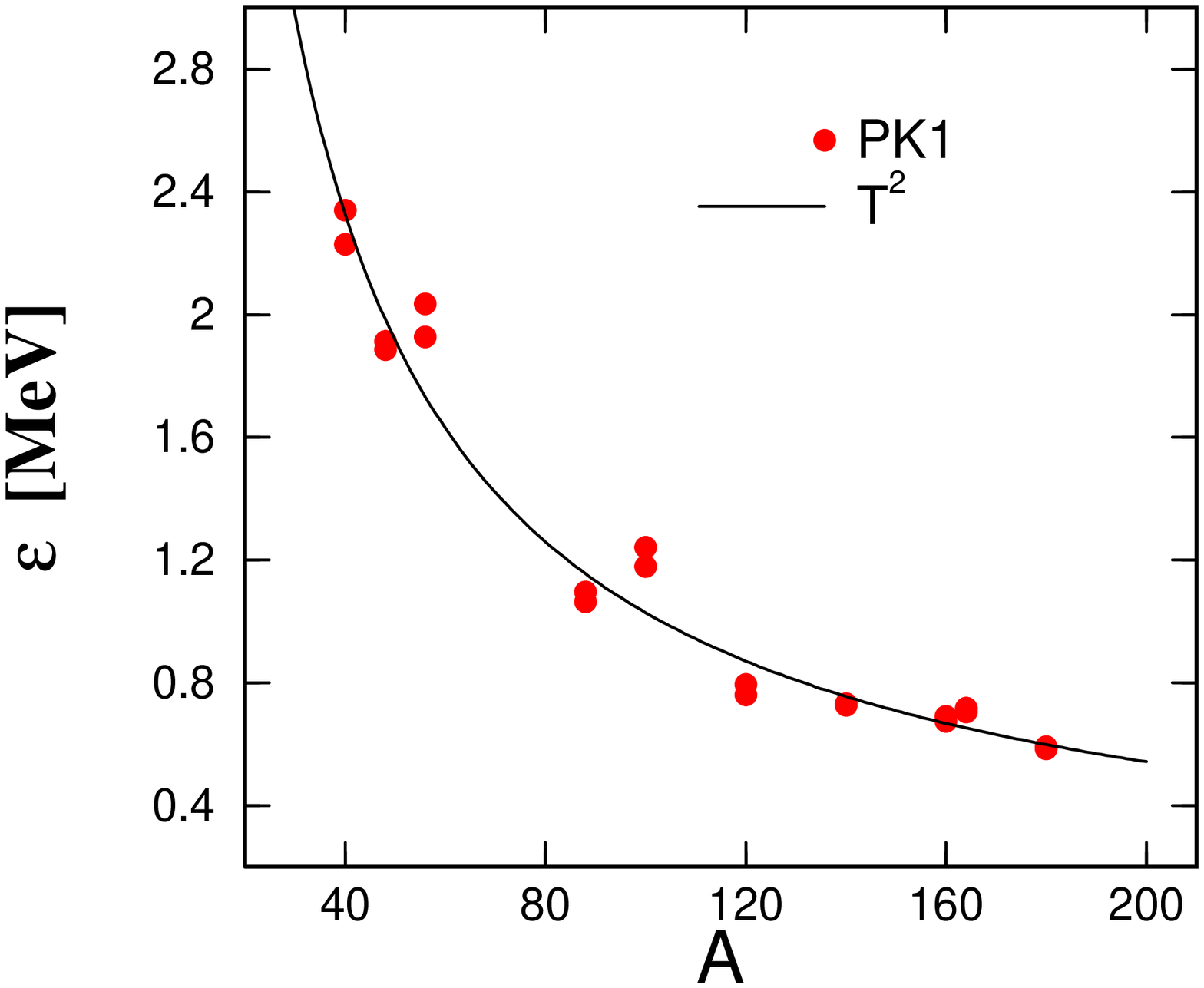}
\includegraphics[width=6.0cm,angle=-90]{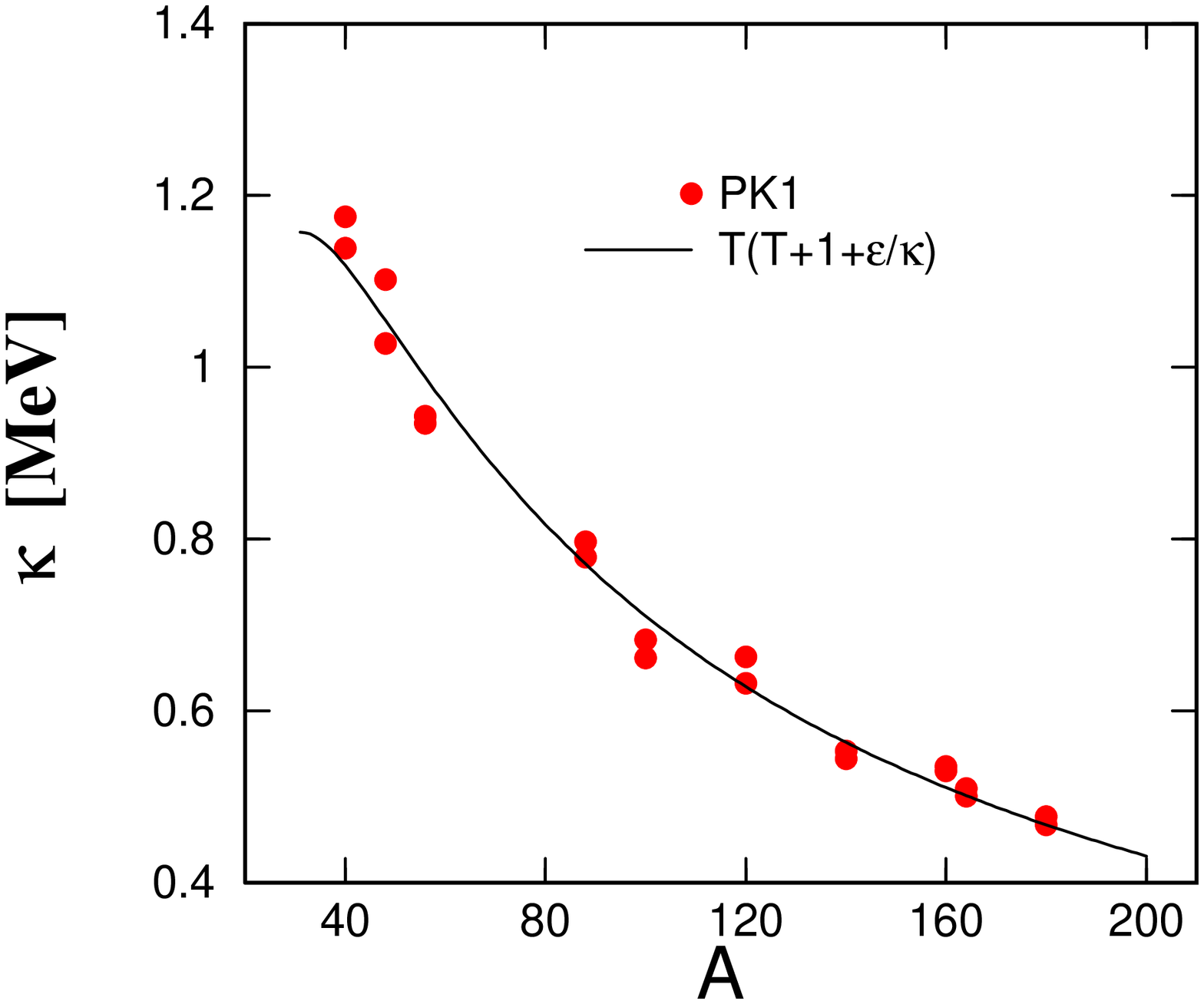}
\caption{\label{fig4} Volume and surface
          contributions to the mean level spacing  $\varepsilon$ (left) and
          effective strength $\kappa$ (right).
          The fitted line is obtained from
          the same points as in Fig.~\ref{fig3new}, i.e.,
          $\varepsilon = \dfrac {130.42} {A} - \dfrac {127.83} {A^{4/3}}$
and
          $\kappa = \dfrac {144.77} {A} - \dfrac {342.08} {A^{4/3}}$. }
\end{figure}


In a similar manner, the volume and surface contributions to the
mean level spacing $\varepsilon$ and the average effective
strength $\kappa$ are determined from the calculations. In the
left panel of Fig.~\ref{fig4}, the mean level spacing
$\varepsilon$ (filled circles) extracted from $\tilde{E}_T -
\tilde{E}_{T=0}=\dfrac 1 2 \varepsilon T^2$ are presented together
with the smooth curve
 obtained from least-square fit:
\begin{equation}
      \varepsilon = \dfrac {130.42} {A} -
                    \dfrac {127.83} {A^{4/3}} ~\text{[MeV]}.
      \label{varepsilonF}
\end{equation}
 In the right panel of Fig. \ref{fig4}, the average effective
strength $\kappa$ (filled circles) extracted from ${E}_T -
\tilde{E}_{T}=\dfrac 1 2 \kappa T(T+1+\varepsilon/\kappa)$ are
presented together with the smooth curve obtained by least-square
fit:
\begin{equation}
      \kappa = \dfrac {144.77} {A} -
               \dfrac {342.08} {A^{4/3}} ~\text{[MeV]}.
      \label{varepsilonK}
\end{equation}
It should be noted that we choose the same nuclei for the fit of
$\varepsilon$, $\kappa$, and $a_\text{sym}$ in the least-square
fitting.

\vspace{0.4cm}


\begin{figure}[tbp]
\includegraphics[width=8cm,angle=-90]{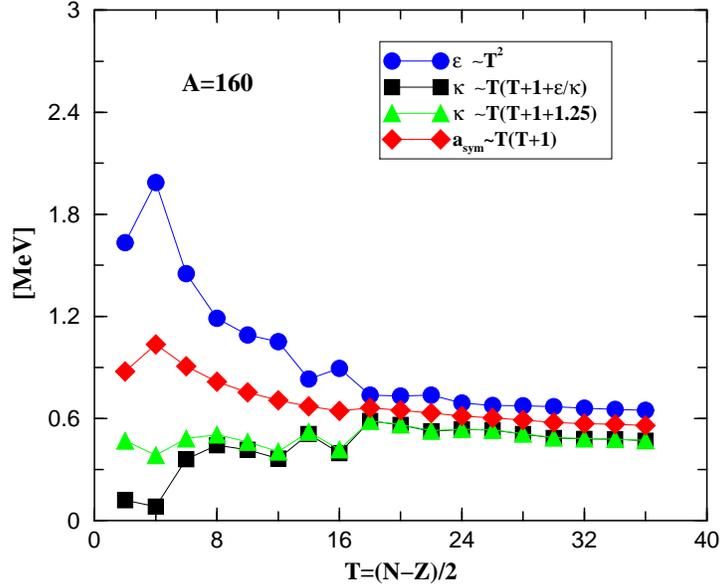}
\caption{\label{fig5} The mean level spacing $\varepsilon$, the
          average effective isovector potential strength $\kappa$
          calculated assuming
          either $T(T+1+\varepsilon/\kappa)$ or $T(T+1+1.25)$
          dependence, and the
          symmetry energy coefficient $a_{sym}$ for the $A=160$ isobaric chain
          in RMF theory with effective Lagrangian PK1. The value of 1.25 is
          the average for $\varepsilon/\kappa$ at $A=160$.}
\end{figure}

 It was shown before that for relatively large values of $T_z$ both
 $\varepsilon (A,T_z)\approx \varepsilon (A)$ and $\kappa (A,T_z)\approx
 \kappa(A)$ along each isobaric chain.
 Let us now fix the value of $A$ and investigate fine effects
 in  $\varepsilon (A,T_z)$, $\kappa (A,T_z)$, and
 $a_{sym}(A,T_z)=\dfrac{1}{2} (\varepsilon (A,T_z)+\kappa (A,T_z))$
  versus $T_z$ in order to study the response of the isovector
potential to changes in shell structure which are naturally
incorporated in the mean level spacing  $\varepsilon (A,T_z)$.
Therefore, we present in Fig.~\ref{fig5} the values of
 $\varepsilon (A,T_z)$, $\kappa (A,T_z)$, and
$a_{sym}(A,T_z)$ calculated using $T^2$,
$T(T+1+\varepsilon/\kappa)$, and  $T(T+1)$  isospin dependencies
 respectively, for the $A=160$ isobaric chain using the parametrization
PK1 of the effective Lagrangian. To avoid the direct connection
between $\kappa$ and $\varepsilon$ entering the
$T(T+1+\varepsilon/\kappa)$ dependence used to extract $\kappa
(A,T_z)$ we also show values of $\kappa (A,T_z)$ obtained assuming
$T(T+1+1.25)$ dependence, where 1.25 represents the average value
of $\varepsilon/\kappa$. Clearly, there are small variations in
 $\varepsilon (A,T_z)$ for $T_z=12, 14, 16$ and
$22$ due to the shell structure. Whilst the same variations are
also obtained for $\kappa(A,T_z)$ at the same value of $T_z$, the
sum $a_{sym}(A,T_z)$ is very smooth. Apparently, changes in
$\varepsilon(A,T_z)$ and $\kappa(A,T_z)$ cancel to large extent.
We also see that the variations in $\kappa (A,T_z)$ are not due to
the $\varepsilon /\kappa$ term,  since similar variations are
found for the $\kappa (A,T_z)$ curve calculated using the
$T(T+1+1.25)$ dependence, see also the curves for $T^2$  and
$T(T+1)$ as shown in the right panel of Fig.~\ref{fig2}.
Apparently, the variations of $\kappa (A,T_z)$  reflect a direct
response to changes in $\varepsilon (A,T_z)$ showing that
effectively, the isovector potential and isoscalar potential
become closely linked.

 \begin{figure}[tbp]
 \includegraphics[width=5.5cm,angle=-90]{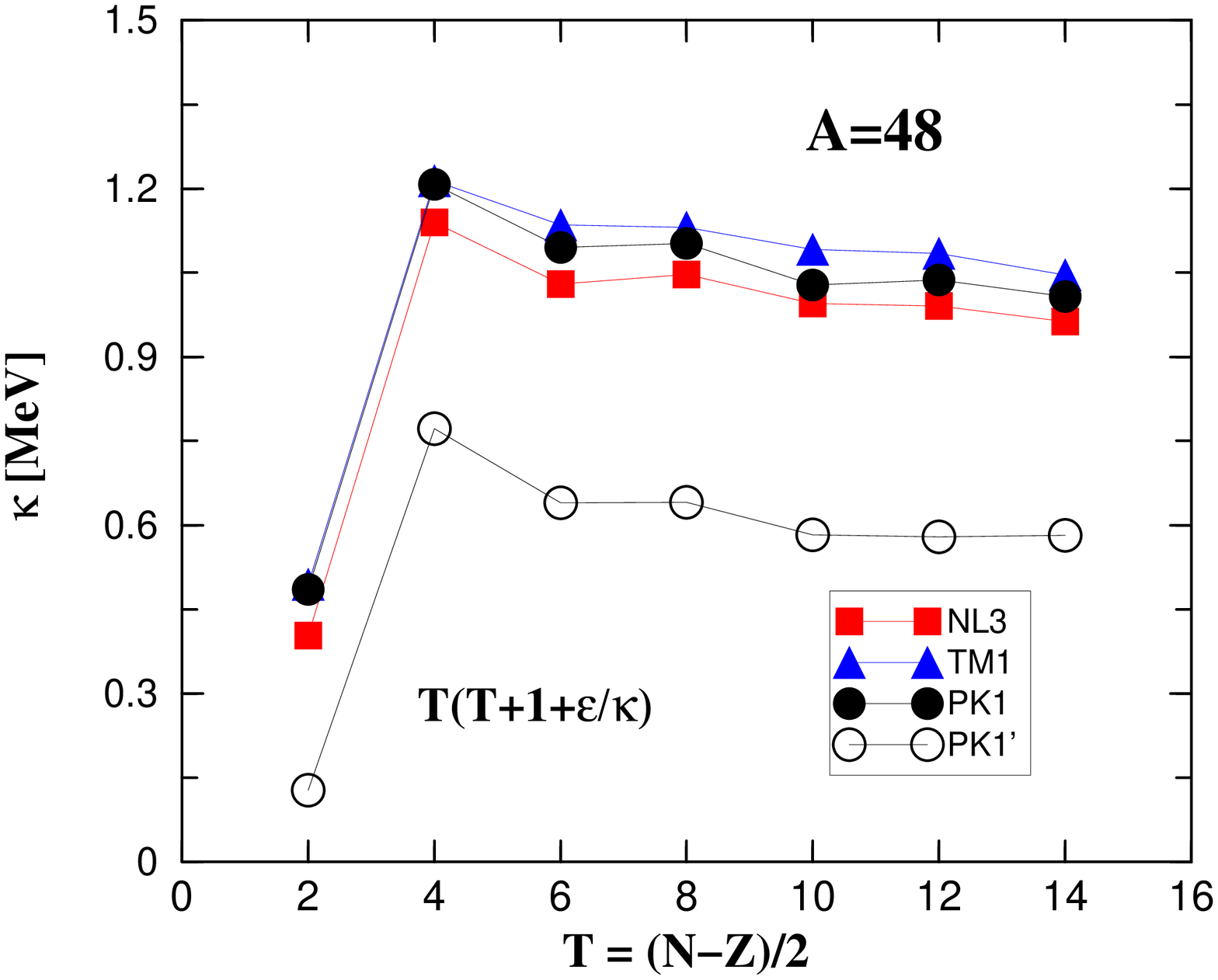}
 \includegraphics[width=5.5cm,angle=-90]{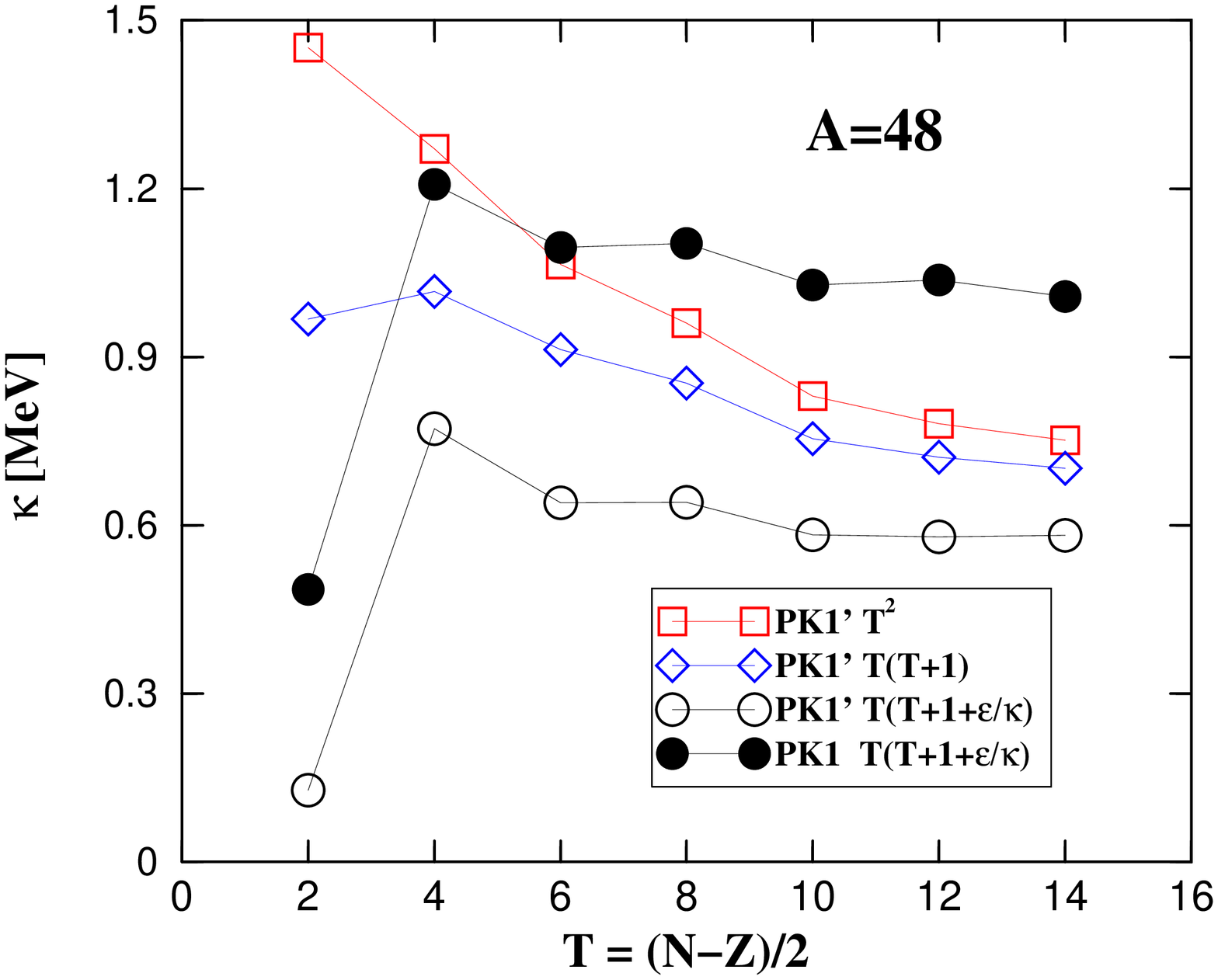}
 \caption{\label{fig6} The comparison of the average effective
          strength $\kappa$ of the isovector potential for the A=48 isobaric chain
          between self-consistent RMF calculation (PK1) and the
          corresponding non self-consistent (PK1'). See the text for further
           explanations. }
 \end{figure}

 In order to understand the origin of the linear term
in RMF theory, self-consistent calculations for the effective
Lagrangian PK1 have been performed including the $\sigma$ and
$\omega$ meson-fields only. With the nucleon densities thus
obtained, the influence of the $\rho$ meson field has been
extracted by switching on the isovector potential in a
non-self-consistent way. The results for the $A$=48 isobaric
chains are labeled as PK1' in Fig.~\ref{fig6}. It is interesting
to note that just by switching on the isovector potential, the
relation $T(T+1+\varepsilon/\kappa)$ is followed quite well,
although it accounts for only $2/3$ of the total $\kappa$. We can
therefore conclude that the linear term $(\varepsilon+\kappa)T$
exists in RMF whenever the isovector potential is present. The
self-consistency between the isovector and isoscalar potentials
roughly contributes to another $1/3$.

\vspace{0.4cm}

As shown in Fig.~\ref{fig3new}, the nuclear symmetry energy for
finite nuclei calculated in RMF theory with PK1 is in good
agreement with the experiment. The symmetry energy coefficient, as
obtained from finite nuclei can be extrapolated to infinite
nuclear matter by setting the surface term of
Eq.~(\ref{EmpSymmEC}) or Eq.~(\ref{SymmEC}) to zero. The
theoretical asymptotic value obtained in this manner $a_{sym}'
\equiv A*a_{sym}/4  \approx 33.6$ MeV is rather close to the
empirically accepted values but considerably smaller [by
$\sim$10\%] than the so called infinite symmetric nuclear
matter(NM)
 values  which are equal to 37.4\,MeV for NL3~\cite{[Lal97]}, 36.9\,MeV
 for TM1~\cite{[Sug95]}, and 37.6 MeV for PK1~\cite{[Lon04]}, respectively.
 The infinite NM values are obtained using the  following formula~\cite{[Gle97]}:
 \begin{equation}
 \label{asym'}
 a_{sym}'=\frac{1}{2}\left(\frac{g_\rho}{m_\rho}\right)\rho_0
         +\frac{k_F^2}{6\sqrt{k_F^2+m^{*2}}},
 \end{equation}
 where $\rho_0$ is the saturation point, $k_F=(3\pi^2\rho_0/2)^{1/3}$
 is the Fermi momentum and $m^*$ is the effective mass at saturation point.
 One should note that the symmetry energy coefficient $a_{sym}'$
 decreases with increasing $N/Z$ ratio in infinite NM. Moreover,
 there are effective Lagrangians having smaller
 values of  $a_{sym}'$ in the infinite symmetric NM including
 GL-97 (32.5\,MeV)~\cite{[Gle97]}, TW-99 (32.77\,MeV)~\cite{[Typ99]} and DD-ME1
 (33.06\,MeV)~\cite{[Nik02]}.
 At present, the origin of this discrepancy is not clear to us.
Certainly, it reflects the important role played by the nuclear
surface in finite nuclei, where one can notice the large
contribution to the surface energy coming from the isovector
potential. Differences in the linear term affect the size of the
surface energy coefficient obtained in the calculations. However,
the volume term of the total symmetry energy coefficient is not
affected much even when varying the linear term between 1 and 4.

\vspace{0.8cm}

In summary, the nuclear symmetry energy has been studied in RMF
theory with effective Lagrangians NL3, TM1, and PK1. The mean
level spacing $\varepsilon$, the effective isovector potential
strength $\kappa$, and the nuclear symmetry energy coefficient
$a_\text{sym}$ are calculated for the isobaric chains $A$=40, 48,
56, 88, 100, 120, 140, 160, 164, and 180 from $T=0$ to the
vicinity of the drip line. It is shown that, except some strong
variations at small values of $T_z$, the mean level spacing
$\varepsilon$ is stabilized at large $T_z$ and lies in the region
of the empirical value after being re-scaled by the effective
mass. These results confirm the general formulation of the
symmetry energy obtained from the simple iso-cranking model as
first proposed in Ref.~\cite{[Sat01x]} and are in agreement with
Skyrme-Hartree-Fock calculations~\cite{[Sat03],[Sat05]}.

By switching on the isovector potential due to the $\rho$ meson
field, the effective isovector potential strength $\kappa$ is
extracted. It is surprising to find that the RMF theory, which is
a Hartree approximation at first glance, generates a large
linear term corresponding at least to ${E_T} - \tilde{E_{T}}
\approx\dfrac 1 2 \kappa (A) T(T+1+\varepsilon/\kappa)$. This is
in contrast to the SHF model, where the isovector potential has a
$\dfrac 1 2 \kappa (A) T(T+1)$ dependence. Hence, the total
nuclear symmetry energy in RMF follows the $T(T+1)$ relation quite
well.

The nuclear symmetry energy coefficient $a_\text{sym}$ as
extracted from $E_\text{sym}=a_\text{sym}T(T+1)$ is in good
agreement with the empirical data in Refs.~\cite{[Jan65x],[Duf95]}.
The discrepancy between the calculated asymptotic value of
$a_{sym}$ and the infinite symmetric nuclear matter estimate
requires further systematic studies. Our work also indicates that
a general formulation of the nuclear binding energy in terms of
the density functional theory in fact can yield a $T(T+1)$
dependence of the symmetry energy. The question of the physical
mechanism leading to the restoration of the complete linear term
within RMF theory is left open.


\begin{acknowledgments}
This work is supported by the Swedish Institute (SI),
the Major State Basic
Research Development Program Under Contract Number G2000077407,
the National Natural Science Foundation of China under Grant No.
10435010, and 10221003, the Doctoral Program Foundation from
the Ministry of Education in China
and by the
Polish Committee for Scientific Research (KBN) under contract
No. 1~P03B~059~27. J. Meng
acknowledges support from STINT and S. Ban would like to thank S.Q.
Zhang for illuminating discussions.
\end{acknowledgments}



\end{document}